\documentclass[epj]{svjour}
\usepackage{graphicx}
\title{Scaling property of the critical hopping parameters for the Bose-Hubbard model}

\begin{document}

\author{Niklas Teichmann\thanks{\email{teichmann@theorie.physik.uni-oldenburg.de}} \
	\and Dennis Hinrichs }

\institute{
 Institut f\"ur Physik - Carl von Ossietzky Universit\"at - D-26111 Oldenburg - Germany\\
}

\abstract{Recently precise results for the boundary between the Mott insulator
	phase and the superfluid phase of the homogeneous Bose-Hubbard model
	have become available for arbitrary integer filling factor $g$ and any
	lattice dimension $d \ge 2$. We use these data for demonstrating that the
	critical hopping parameters obey a scaling relationship which allows
	one to map results for different $g$ onto each other. Unexpectedly,
	the mean-field result captures the dependence of the exact critical
	parameters on the filling factor almost fully. We also present an
	approximation formula which describes the critical parameters
	for $d \ge 2$ and any $g$ with high accuracy.}
	


\maketitle

	In the last ten years experiments with ultracold atoms in optical lattice 
	potentials generated by laser beams have become an important tool for 
	mimicking phenomena in solid state physics, and have contributed to further 
	understanding of many-particle systems~\cite{Lewenstein2007,Bloch2008}. Outstanding 
	examples are the observation of Bloch oscillations~\cite{Salomon1996,Morsch2001}, the 
	direct measurement of Anderson localisation~\cite{Billy2008}, or the detection of the 
	Mott insulator-to-superfluid transition~\cite{GreinerEtAl02}.
	A model of utmost importance in this field of research is the Bose-Hubbard model.
	It describes a system of interacting ultracold bosons in an optical lattice~\cite{JakschEtAl98}, 
	and has been intensively studied since the late eighties with numerous
	methods (see \textit{e.g.}~\cite{Lewenstein2007,Bloch2008} and references therein). 
	Nevertheless, the Bose-Hubbard model still is a topic of active 
	research~\cite{CapogrossoSansone07,KollathEtAl07,SantosPelster08,PolakKopec09,FreericksEtAl09}. 

	Although many techniques have been developed in order to calculate the phase diagram
	of the homogeneous Bose-Hubbard model, the evolution of the phase boundary between the 
	Mott insulator and the superfluid phase with increasing filling factor so far has not 
	been investigated thoroughly. 
	For example, filling factors greater than one typically occur in experiments with a 
	superimposed harmonic trap, which is necessary for confinement reasons. Then a wedding 
	cake structure emerges with decreasing filling from the trap centre to the outer region.	
	High-order perturbation theory utilising a process-chain
	approach~\cite{Eckardt09} has made the regime of larger filling accessible only very recently, 
	and has provided accurate data of the critical parameters for arbitrary integer filling factors.
	
	In this communication we report a general scaling property of the critical hopping parameters 
	with increasing filling factor. This scaling renders the critical parameters (almost) 
	independent of the filling. Moreover, we present an approximation formula for the critical 
	hopping for any dimension $d\ge 2$, and any filling factor.
	We proceed as follows:
	First the Bose-Hubbard model is briefly recapitulated. We then compare the 
	phase boundary obtained via the process-chain approach~\cite{TeichmannEtAl2009} for unit 
	filling ($g=1$) with two recent high-precision results~\cite{CapogrossoSansone07,FreericksEtAl09}. 
	The excellent agreement serves to underline the high accuracy of the process-chain approach, which 
	also provides data for higher filling factors, for which no reference data are available.	
	After that, we apply three successively more accurate scalings to the critical parameters 
	belonging to different filling factors.	Finally, we develop the approximation formula, and 
	end the letter with some concluding remarks.
	
\section{The Bose-Hubbard model} 
	
	Ultracold bosons in an optical lattice are fairly well described by the
	Bose-Hubbard model as long as the lattice is sufficiently deep~\cite{JakschEtAl98}. 
	In the limit of zero temperature only the lowest Wannier state at each individual 
	site has to be considered, leading to a 
	single-band model. Furthermore the on-site approximation is applied, \textit{i.e.} it is 
	assumed that only particles sitting on the same lattice site interact with each other. 
	This interatomic repulsion leads to an increase of energy by $U$ for each pair of 
	interacting particles. Restricting the hopping processes to adjacent sites only, 
	and denoting the matrix element connecting
	two neighbouring sites by $J$, one arrives at the well-known Bose-Hubbard 
	Hamiltonian (see \textit{e.g.}~\cite{JakschEtAl98}):
	\begin{equation}
		H = - J\sum_{\langle i,j \rangle} \hat{a}_i^{\dagger} \hat{a}_j^{\phantom \dagger}
		+\frac{U}{2} \sum_{i} \hat{n}_i(\hat{n}_i-1) 
		- \mu \sum_{i} \hat{n}_i \;.
		\label{eq:Hamiltonian}
	\end{equation} 
	The operators $\hat{a}_i^{\dagger}$ and $\hat{a}_i^{\phantom \dagger}$ are the 
	bosonic creation and annihilation operators for site $i$, and 
	$\hat{n}_i = \hat{a}_i^{\dagger} \hat{a}_i^{\phantom \dagger}$ is the corresponding 
	number operator. The subscript $\langle i,j \rangle$ indicates that the sum is
	restricted to nearest neighbours. As we intend to describe a homogeneous system, 
	the chemical potential $\mu$ is constant. The competition between the
	kinetic energy, described by the first term of the Hamiltonian~(\ref{eq:Hamiltonian}),
	and the interaction energy (second term) gives rise to a quantum phase 
	transition from a Mott-insulating to a superfluid phase upon increasing 
	the hopping strength~\cite{Sachdev99}.
	
\begin{figure}
		\centering
		\includegraphics[scale=0.3, angle=-90]{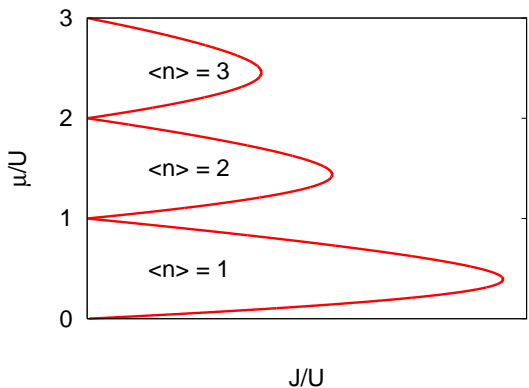}
		\caption{Phase diagram of the Bose-Hubbard model, showing the three lowest Mott lobes.
			Inside the lobes the occupation number $\langle \hat{n} \rangle$ is fixed.}
		\label{fig:phase}	
	\end{figure}

	When using the pair interaction energy $U$ as the energy scale of reference 
	(\textit{i.e.} when dividing eq.~(\ref{eq:Hamiltonian}) by $U$) the Hamiltonian is governed by 
	two dimensionless parameters: The hopping parameter $J/U$ and the scaled chemical
	potential $\mu/U$. In case of $J/U \ll 1$ the system is in a Mott phase, in which particles tend 
	to localise at the individual sites. A fixed number of atoms occupies each lattice site 
	and thus the system is incompressible: $\partial \langle n_i\rangle/\partial \mu = 0$. 
	On the other hand, when $J/U \gg 1$, the particles tend to delocalise, and phase
	fluctuations are suppressed. The bosons then condense into the lowest Bloch state and 
	the system is superfluid. Figure~\ref{fig:phase} sketches the phase diagram in 
	the $J/U$-$\mu/U$-plane. Inside the Mott lobes the occupation number 
	$\langle \hat{n} \rangle$ is integer, while outside the lobes -- in the superfluid phase -- 
	it can take on any rational value.
	The lobe's width at a fixed tunneling rate corresponds to the energy gap for a 
	particle-hole excitation, which represents an important experimentally accessible quantity. 
	It vanishes when the hopping strength exceeds the critical value $(J/U)_{\rm c}$.
	For integer $\mu/U$ and $J/U=0$ the system's ground state is highly degenerate, as
	either $\mu/U$ or $\mu/U+1$ particles can occupy each site.	
	Qualitatively the phase diagram has already been understood by 
	Fisher~\emph{et al.}~\cite{FisherEtAl89}. The mean-field phase boundary, given by
	\begin{equation}
		(J/U)_{\rm pb} = \frac{1}{2d}\frac{(\mu/U-g+1)(g-\mu/U)}{\mu/U+1}
		\label{eq:mf_boundary}
		\;,
	\end{equation} 
	was derived within a model which embodies equal couplings between all sites.
	Here $d$ is the dimensionality of the lattice and $g$ is the integer number 
	of particles per site, referred to as the filling factor of the optical lattice. 
	Precise quantitative results for the phase boundary of the three-dimensional 
	homogeneous Bose-Hubbard model have been obtained by Quantum Monte Carlo
	simulations~\cite{CapogrossoSansone07}, and by means of a strong coupling 
	expansion together with a scaling analysis~\cite{FreericksEtAl09}, for $g=1$. 
	So far, a precise analysis of the behaviour of the phase boundary with increasing 
	filling factor has not been performed. It should be pointed out that accurate data 
	for the phase boundary are hard to obtain at higher fillings, as the dimensionality 
	of the Hilbert space rapidly grows with increasing number of particles.

\section{Comparison of approaches for unit filling} 
	
\begin{figure}
		\includegraphics[scale=0.3,angle=-90]{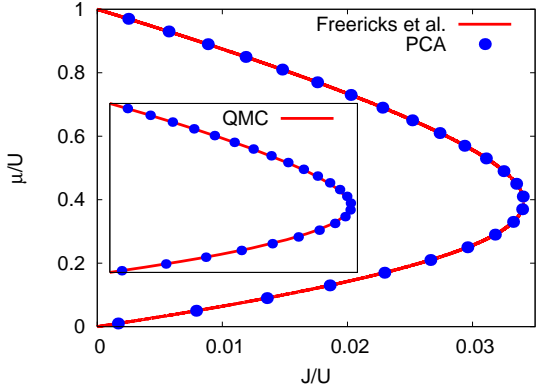}
		\caption{Comparison of phase boundaries for $d=3$ and unit filling ($g=1$) 
			obtained from the process-chain approach (PCA, ref.~\cite{TeichmannEtAl2009}), 
			and from the scaling approach (Freericks~\emph{et al.}, ref.~\cite{FreericksEtAl09}). 
			The inset includes data from Quantum Monte Carlo
			simulations (QMC, ref.~\cite{CapogrossoSansone07}). 
			Especially the results by Freericks~\emph{et al.}\ are virtually 
			indistinguishable from our findings, while the QMC data show exiguous
			deviations close to the tip~\cite{Teichmann_arxiv}.
			}
		\label{fig:comparison}
	\end{figure}

	Recently dos~Santos \emph{et al.}\ derived the phase 
	boundary for the Bose-Hubbard model by applying the method of the effective 
	potential~\cite{SantosPelster08}. Employing the process-chain approach~\cite{Eckardt09} 
	together with the effective potential method, high orders in the hopping parameter $J/U$  
	have become accessible, resulting in precise phase boundaries for dimensions $d\geq 2$ 
	and arbitrary integer filling factors~\cite{TeichmannEtAl2009}.	Figure~\ref{fig:comparison} 
	displays a comparison between the boundaries for $g=1$ obtained within three different methods: 
	the process-chain approach (PCA), a strong coupling expansion paired with a scaling
	analysis~\cite{FreericksEtAl09}, and Quantum Monte Carlo simulations~\cite{CapogrossoSansone07}. 
	The latter two techniques are known to be highly accurate, so that the perfect agreement proves the
	reliability of the PCA data for unit filling. In the following we rely on the fact that the 
	process-chain approach reaches comparable accuracy for any $g$.
%

\section{Scaling behaviour with increasing filling} 

	In order to monitor the evolution of the phase boundary with increasing filling factor 
	$g$ we concentrate on the critical value $(J/U)_{\rm c}$, which is defined by the maximum 
	hopping	parameter of the respective Mott lobe. All outcomes
	reported here are obtained by a high-order perturbation calculation up to eighth order 
	in $J/U$, with an additional extrapolation to infinite order. For a detailed explanation
	how to apply the process-chain approach to the Bose-Hubbard model we refer to
	ref.~\cite{Teichmann_arxiv}. 
	Although the data employed here were calculated for a hypercubic lattice, one can apply 
	the same technique to other types of lattices such as triangular ones or ladder systems.
	
\begin{figure}			
		\includegraphics[scale=0.3,angle=-90]{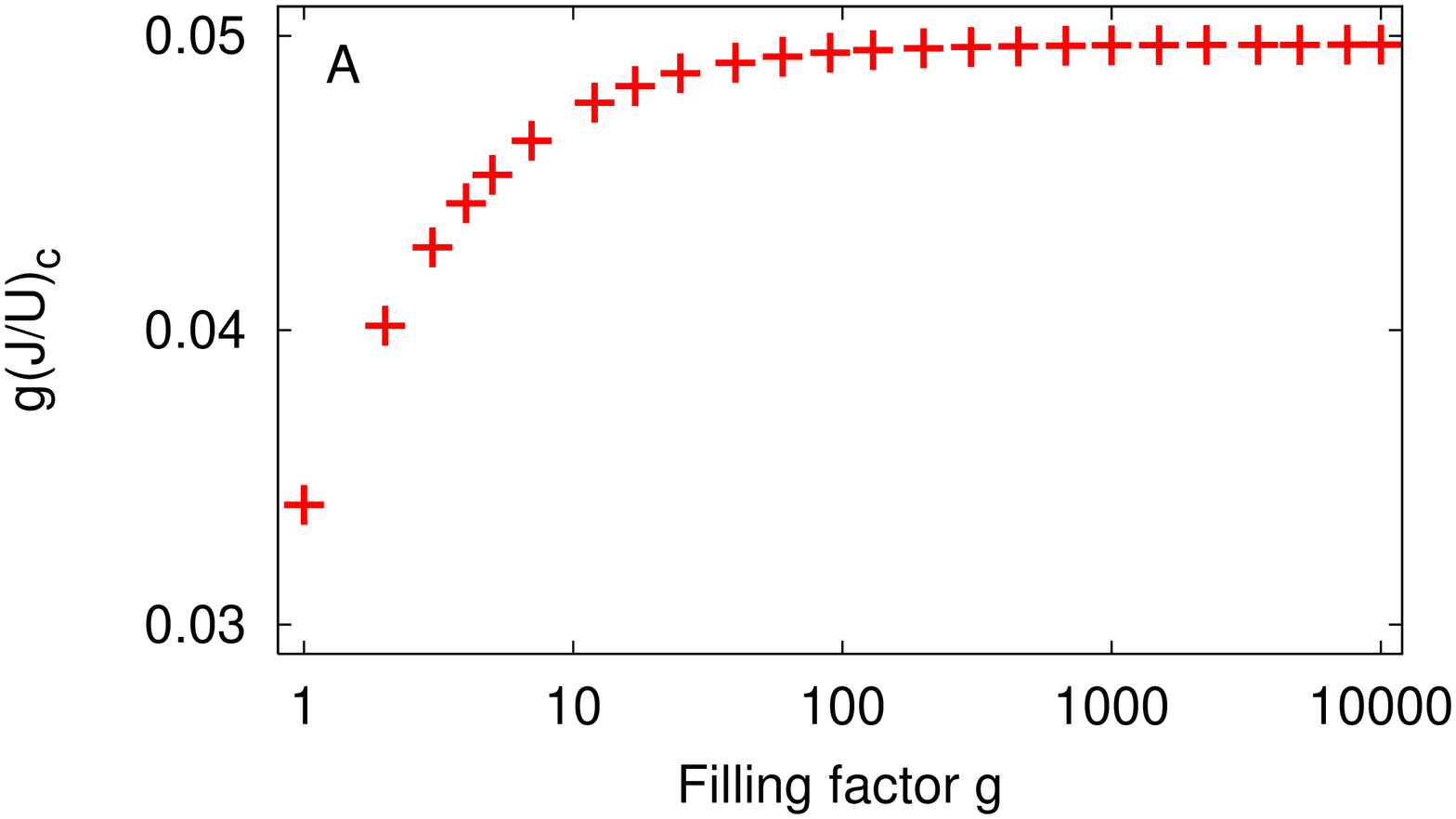}
		\includegraphics[scale=0.3,angle=-90]{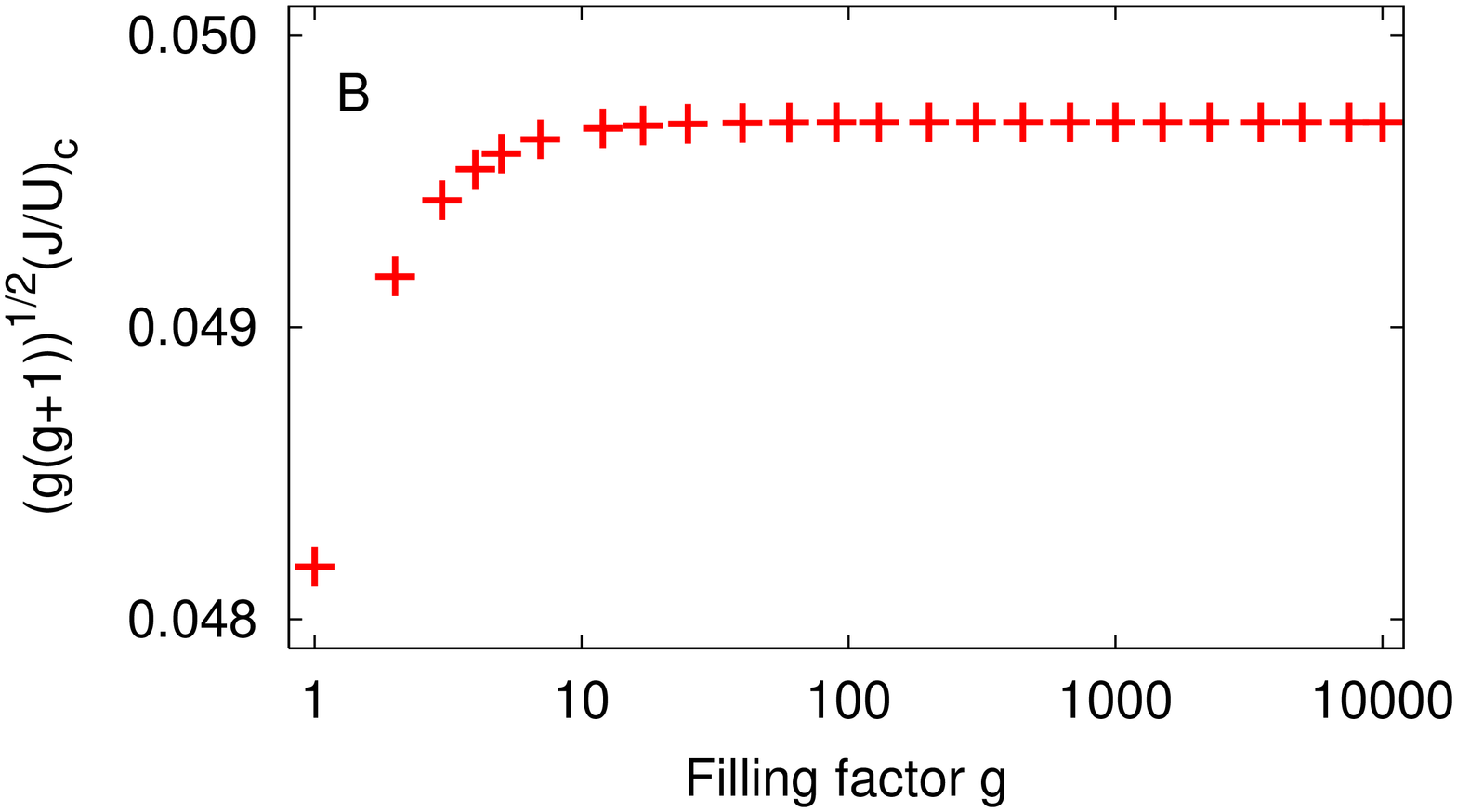}
		\caption{Critical hopping parameters $(J/U)_{\rm c}$ for various
			filling factors $g$ in three dimensions and two different 
			kinds of scaling.
			A: The values are scaled by the factor $g$.
			B: The values are scaled by the factor $\sqrt{g(g+1)}$.
			Note the finer scale in the lower figure.}
		\label{fig:Jc_g12}
	\end{figure}

	The usual expectation that $(J/U)_{\rm c}$ scales like $\sim 1/g$ for high fillings $g$ 
	is borne out by the upper plot in fig.~\ref{fig:Jc_g12}, since the 
	scaled critical hopping
	\begin{equation}
		\widetilde{\left(\frac{J}{U}\right)}_{\rm c} = 
		g\left(\frac{J}{U}\right)_{\rm c}\;,
		\label{eq:scal1}
	\end{equation}
	approaches a constant for large $g$. But these scaled hopping parameters still
	cover a broad range (from $0.034$ to $0.05$), amounting to a variation of 32\%
	and thus indicating that this type of scaling might be suboptimal.	 
	Indeed, a considerably better picture is obtained when employing the phenomenological 
	scaling 
	\begin{equation}
		\widetilde{\left(\frac{J}{U}\right)}_{\rm c} = 
		\sqrt{g(g+1)}\left(\frac{J}{U}\right)_{\rm c}\;,
		\label{eq:scal2}
	\end{equation}
	which was discovered in ref.~\cite{Teichmann_arxiv} for energy corrections,
	and which stems from the factors gained by successive hopping processes 
	on a lattice with $g$ particles per site on average. As the lower plot in 
	fig.~\ref{fig:Jc_g12} shows, the improvement over the simple ansatz~(\ref{eq:scal1}) 
	displayed in the upper plot of fig.~\ref{fig:Jc_g12} corresponds to a factor of $10$: 
	The scaled critical parameters now range only from $0.048$ to $0.050$, amounting 
	to a variation of 3.1\% . Of course, in the limit of high filling factors both scalings 
	yield the same results, as then $\sqrt{g(g+1)} \approx g$.

\begin{figure}
		\includegraphics[scale=0.3,angle=-90]{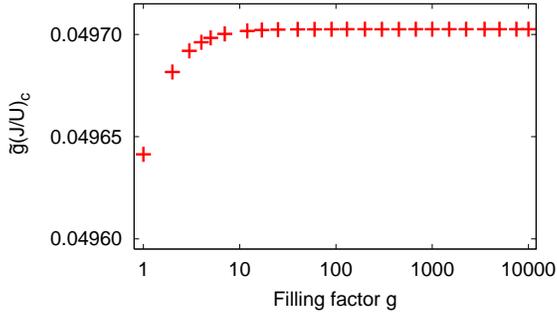}
		\caption{Critical hopping parameters $(J/U)_{\rm c}$ for various
		filling factors $g$ in three dimensions. The values are scaled by the factor
 		$\widetilde{g}=\sqrt{g(g+1)}\left(\frac{1}{2}+\sqrt{\frac{1}{4}+\frac{1}{16g(g+1)}}\right)$,	
		which emerges in the mean-field limit. Observe the fairly fine scale.}
		\label{fig:Jc_g3}
	\end{figure}

	Although eq.~(\ref{eq:scal2}) already leads to satisfying results,
	a further improvement can be achieved by examining
	the mean-field phase boundary of the Bose-Hubbard model, which is exact in 
	the limiting case of infinite dimensionality (see \textit{e.g.}~\cite{TeichmannEtAl2009}). 
	By finding the maxima of eq.~(\ref{eq:mf_boundary}) with respect to the scaled chemical 
	potential $\mu/U$, the critical mean-field hopping parameters are identified as
	\begin{equation}
		\left(\frac{J}{U}\right)^{\rm MF}_{\rm c} =\frac{1}{2d} 
		\left(2g+1 - 2\sqrt{g(g+1)} \right)
		\;.
		\label{eq:JU_c_mf}
	\end{equation} 
	Under the hypothesis that this dependence on the filling factor $g$ also 
	describes the $g$-dependence for finite dimensionality, we attempt the scaling	
	\begin{equation}
		\widetilde{\left(\frac{J}{U}\right)}_{\rm c} = 
		\frac{1}{4}\left(2g+1 - 2\sqrt{g(g+1)} \right)^{-1}
		\left(\frac{J}{U}\right)_{\rm c}
		\label{eq:scal3}
		\;,
	\end{equation} 
	where the factor of $1/4$ has been included in order to ensure that the entire
	prefactor correctly reduces to $g$ for high filling.
	As witnessed by fig.~\ref{fig:Jc_g3} for $d=3$, this scaling works astonishingly well:
	The scaled parameters depicted there vary by merely 0.12\%. In view of the accuracy of the 
	process-chain approach itself, which is about 1\% for $(J/U)_{\rm c}$ in
	a three-dimensional system~\cite{TeichmannEtAl2009}, these scaled critical parameters~(\ref{eq:scal3}) 
	can be regarded as constant. For other dimensions $d$ we find the same marvellous mapping, 
	making the critical hopping strengths practically independent of the filling factor $g$.
	
	One can rewrite the ansatz~(\ref{eq:scal3}) and finds after some lines of algebra
	\begin{equation}
		\widetilde{\left(\frac{J}{U}\right)}_{\rm c} = 
		\sqrt{g(g+1)}\left[\frac{1}{2}+\sqrt{\frac{1}{4}+\frac{1}{16g(g+1)}} \,\right]
		\left(\frac{J}{U}\right)_{\rm c}
		\label{eq:scal3b}
		\;,
	\end{equation}
	which reveals that this scaling covers the two previous attempts of 
	eqs.~(\ref{eq:scal1}) and~(\ref{eq:scal2}) in the limit of high filling 
	factors~$g$.
	Hence we conclude that for any $d$ the functional dependence on the filling factor 
	is correctly reflected by the scaling~(\ref{eq:scal3b}), which was extracted from the 
	mean-field result~(\ref{eq:JU_c_mf}). This appears to be unexpected, and could
	not be validated before reliable data for large $g$ have been available.
	Of course, deviations of $(J/U)_{\rm c}$ itself from the mean-field result due to 
	finite dimensionality remain.	

\section{An approximation formula for the critical hopping} 
		
\begin{figure}
		\includegraphics[scale=0.3,angle=-90]{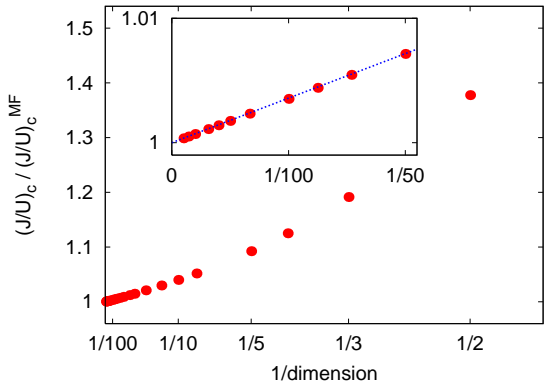}
		\caption{Critical hopping parameter $(J/U)_{\rm c}$ normalised by the 
			mean-field value $(J/U)^{\rm MF}_{\rm c}$ plotted versus $1/d$ for 
			unit filling. The inset shows a magnification for high $d$ and
			includes a linear fit $f(1/d) = 1+0.358/d$ indicated by the dotted line.}
		\label{fig:Jc_d}
	\end{figure}

	With the process-chain approach one is also able to monitor the convergence of the 
	critical hopping parameter towards the mean-field prediction~(\ref{eq:JU_c_mf}) with 
	increasing lattice dimensionality $d$. When plotting the relative values
	$(J/U)_{\rm c}/(J/U)^{\rm MF}_{\rm c}$ of the critical parameters versus $1/d$, 
	this convergence can be nicely watched. Figure~\ref{fig:Jc_d} shows that even for 
	quite high dimensionalities $d$ of the lattice the mean-field theory substantially 
	overestimates the critical hopping, for instance, by 4\% for $d=10$. The deviations 
	in two and three dimensions are as large as 38\% and 19\%, respectively.

	In order to set up an approximation formula for the critical hopping parameter for
	every dimension and arbitrary filling factor, one can now focus on the 
	dimension-dependence since the dependence on the filling factor $g$ is already
	accurately described by the scaling~(\ref{eq:scal3b}). We therefore perform our 
	investigations on the data for unit filling. 
	A reasonable ansatz for $(J/U)_{\rm c}(d,g)$ is an expansion in $1/d$:
	\begin{equation}
		\left(\frac{J}{U}\right)_{\rm c}(d,g) = 
		\left(\frac{J}{U}\right)^{\rm MF}_{\rm c}
		\left[1+\sum_{\ell=1}^N a_{\ell} \frac{1}{d^\ell} \right]
		\;.
		\label{eq:expansion}
	\end{equation} 
	Indeed we find a mainly linear dependence on $1/d$ for high dimensions $d$, as shown
	by the inset of fig.~\ref{fig:Jc_d}. Fixing $N=3$, fitting the 
	expansion~(\ref{eq:expansion}) to the data points given in fig.~\ref{fig:Jc_d},
	and inserting $(J/U)^{\rm MF}_{\rm c}$, one obtains
	\begin{eqnarray}				
		\left(\frac{J}{U}\right)_{\rm c}(d,g)
		&=&\frac{1}{2d} \left(2g+1 - 2\sqrt{g(g+1)} \right)
		 \label{eq:approx}\\ 
		&\times& 
		\left[1+0.35\frac{1}{d}+0.39\frac{1}{d^2} + 0.84\frac{1}{d^3} \right]
		\nonumber \;.		
	\end{eqnarray} 
	This formula approximates the critical values $(J/U)_{\rm c}$ obtained within the framework of the 
	process-chain approach with an accuracy of better than $0.15$\%. Up to first order in $1/d$
	findings deducible from ref.~\cite{SantosPelster08} are in very good agreement with this result.

\section{Conclusion} 

	When employing the method of the effective potential combined with the 
	process-chain approach, one obtains high-precision data for the critical 
	hopping parameter $(J/U)_{\rm c}$ for any lattice dimension $d\geq 2$ 
	and arbitrary integer filling factor $g$. These values show excellent 
	agreement with other recent findings for the case $d=3$ and 
	$g=1$~\cite{CapogrossoSansone07,FreericksEtAl09}, thus 
	gauging the accuracy of the data. We have shown that the relatively simple 
	scaling~(\ref{eq:scal3b}), which was taken over from the mean-field result, 
	maps the data for different filling factors almost perfectly onto each other. 
	Hence the dependence of $(J/U)_{\rm c}$ on the filling $g$ is known with 
	high precision. Although not theoretically deduced but accurately fitted, 
	we have presented an excellent approximation formula [eq.~(\ref{eq:approx})], 
	which provides $(J/U)_{\rm c}$ for any dimensionality $d\ge2$ and any filling 
	factor $g$ for the homogeneous Bose-Hubbard model. These results may serve 
	as benchmarks for future developments of other methods.
	
\begin{acknowledgement}
	We thank M.~Holthaus for valuable comments and careful reading 
	of the manuscript, and B.~Capogrosso-Sansone for providing the QMC data. 
	Computer power was obtained from the GOLEM~I 
	cluster of the Universit\"at Oldenburg. N.T.\ acknowledges a fellowship
	from the Studienstiftung des deutschen Volkes. Furthermore, financial 
	support by the Deutsche Forschungsgemeinschaft (DFG) is gratefully 
	acknowledged.
\end{acknowledgement}

 
\end{document}